\documentclass[nofootinbib,aps,11pt,letterpaper,superscriptaddress]{revtex4}
\usepackage{graphicx}
\usepackage{hyperref}
\usepackage{cancel}
\usepackage{amssymb}
\usepackage{textcomp}
\usepackage{amsmath}
\usepackage{bm}
\usepackage{times}
\usepackage{epsfig}
\usepackage{color}
\usepackage{slashed}

\newcommand{\MeV}{{\rm MeV}}

\usepackage{ulem,fancyvrb}
\usepackage{xcolor}

\begin{document}
\title{\LARGE  Neutrino dipole portal at electron colliders}
\bigskip
\author{Yu Zhang}
\affiliation{School of Physics, Hefei University of Technology, Hefei 230601,China}
\author{Mao Song}
\affiliation{School of Physics and Optoelectronics Engineering, Anhui University, Hefei 230601,China}
\author{Ran Ding}
\affiliation{School of Physics and Optoelectronics Engineering, Anhui University, Hefei 230601,China}
\author{Liangwen Chen}
\email{chenlw@impcas.ac.cn}
\affiliation{Institute of Modern Physics, CAS, Lanzhou 730000, China}
\affiliation{Advanced Energy Science and Technology Guangdong Laboratory, Huizhou 516000, China}

\begin{abstract}
We propose to search for  a heavy neutral lepton (HNL), that is also know 
as sterile neutrino, in electron colliders running with the center-of-mass energies
at few GeV, including BESIII, Belle II, and the proposed Super Tau Charm Factory (STCF). 
We consider the HNL interacting with Standard Model neutrino and  photon via a transition magnetic moment, the so-called dipole portal.
We use the monophoton signature at electron colliders to probe the 
constraints on the active-sterile neutrino transition magnetic moments $d$ as the function of the HNL's mass $m_N$.
It is found that BESIII, Belle II and STCF 
can probe the upper limits for $d$ down to 1.3 $\times 10^{-5}\ {\rm GeV}^{-1}$, 8 $\times 10^{-6}\ {\rm GeV}^{-1}$, and 1.3 $\times 10^{-6}\ {\rm GeV}^{-1}$ with $m_N$ around GeV scale, respectively, and have sensitivity to the previously unexplored parameter space for electron- ($d_e$) and tau-neutrino ($d_\tau$) dipole portal with $m_N$ from dozens to thounsands MeV. 
On  $d_\mu$ for HNL mixing with the {muon}-neutrino, Belle II and STCF can also provide leading constraints. 

\end{abstract}

\maketitle

\section{Introduction}
The discovery of neutrino flavor oscillations confirms 
{the existence of neutrino mass and mixing }.
This fact is one of the key observational facts indicating 
the imperfection of Standard Model (SM).
Models containing new neutral fermionic states $N$, which can
connect with neutrino masses via the so-called neutrino portal interaction,
${\cal L}\supset NHL$, have attracted significant attentions in the last few years.
Here, $N$, so-called sterile neutrinos, is singlet under the SM gauge groups and often refered to as heavy neutral leptons (HNL), $L$ is the SM leptons, and $H$ is the Higgs doublet.
The HNLs can mix with the left-handed neutrinos after the Higgs acquires
a vacuum expectation value.

Alternatively, the heavy state $N$ may couple to the SM through a higher-dimensional operator.
In this work, we  consider  a HNL interacting with the SM
via a transition magnetic moment, the so-called  ``neutrino dipole
portal", described by the following term in the Lagrangian:
\begin{equation}
\mathcal{L} \supset d_k \bar{\nu}_{L}^k \sigma_{\mu \nu} F^{\mu \nu} N+\mathrm{H.c.},
\label{eq:L}
\end{equation}
where $k=e,\mu,\tau$ denotes the flavor index of lepton,
$\nu_L$ is a SM left-handed (active) neutrino field,
$\sigma_{\mu \nu}=\frac{i}{2}[\gamma_\mu,\gamma_\nu]$, 
$F^{\mu \nu}$ is the electromagnetic field strength tensor, and $d$ is the active-sterile neutrino transition magnetic moment, that controls the strength of the interaction with the units of (mass)$^{-1}$. 

A lot of attentions have been paid to the neutrino dipole portal,
in the context of various laboratory, astrophysical, and cosmological
{bounds on $d_k$}.
A summary of existing constraints can be found in Refs.\cite{Magill:2018jla,Schwetz:2020xra}, 
including 
 Borexino \cite{Borexino:2017fbd, Brdar:2020quo}, Xenon-1T \cite{XENON:2020rca,Brdar:2020quo}, CHARM-II \cite{CHARM-II:1989srx,Coloma:2017ppo}, MiniBooNE \cite{MiniBooNE:2007uho, Magill:2018jla}, LSND \cite{LSND:1996ubh, Magill:2018jla}, NOMAD \cite{NOMAD:1998pxi,Gninenko:1998nn, Magill:2018jla},  DONUT \cite{DONUT:2001zvi}, LEP \cite{Magill:2018jla}, and  supernovae SN 1987A \cite{Magill:2018jla}, BBN $^4$He abundance \cite{Brdar:2020quo}, solar neutrinos \cite{Brdar:2020quo, Plestid:2020vqf}. 
 The sesitivities of future projects or estimated
 exclusions can also be obtained, such as 
 SHiP \cite{Magill:2018jla}, Forward LHC Detectors \cite{Jodlowski:2020vhr, Ismail:2021dyp}, Icecube  \cite{Coloma:2017ppo}, SuperCDMS \cite{Shoemaker:2018vii}, DUNE \cite{Schwetz:2020xra}, CE$\nu$NS and E$\nu$ES \cite{Miranda:2021kre}.
 In this paper, we investigate the experimental sensitivity on active-sterile neutrino transition magnetic moments from electron colliders operated at the GeV scale, including BESIII \cite{Asner:2008nq}, Belle II \cite{Belle-II:2010dht}, and the proposed Super Tau Charm Factory (STCF) \cite{Luo:2019xqt, Charm-TauFactory:2013cnj,Shi:2020nrf}. 
 
 The  rest paper is organized as follows. In Sec. \ref{sec:sb} we describe the relevant signal and backgrounds at electron colliders. In Sec. \ref{sec:res} we summarize current constraints coming from exsisting experiments, and expected sensitivity on  neutrino dipole coupling
 at BESIII, STCF and Belle II. Finally, a short summary is given is Sec. \ref{sec:sum}.

\section{Electron collider signals}
\label{sec:sb}
In this work, we investigate the Dirac sterible {neutrino} $N$ production via dipole portal at
$e^+e^-$ colliders that are  operated with the center-of-mass
(CM ) energies of several GeV, such as BESIII, Belle II and future STCF.
At these colliders, production will proceed via
$e^+e^-\to \gamma^*\to N\bar\nu$
and $e^+e^-\to \gamma^*\to \bar N\nu$.
The differential cross section for on-shell $N$ and a SM neutrino 
production $e^+e^-\to  N\bar\nu$ is 
\begin{equation}
\frac{d\sigma_{N\bar\nu}}{dz_N}=\frac{d^2 \alpha \left({s-m_N}^2\right)^2 \left((1-z_N^2)s+\left(1+z_N^2\right) m_N^2 \right)}{4  s^3},
\end{equation}
where $\alpha$ is the fine structure constant, $z_N\equiv\cos\theta_N$,
with $\theta_N$ being the relative angle between electron beam axis and 
the momentum of $N$ in the CM frame, $s$ is the square of the CM energy, and $m_N$ denotes the mass of $N$.
{Notice that the production rates of $N$ associated with different flavor neutrino at electron collider are same, thus we omit the lepton  flavor index of $d$ here.  The electron collider sensitivities on transition magnetic moments derived in this letter are equally applicable to all lepton flavor.}
The total production cross section for 
$e^+e^-\to N \bar\nu$
after integrating over 
all angles is 
\begin{eqnarray}
\sigma(e^+e^-\to N \bar\nu)&=&\frac{\alpha d^2 (s-m_N^2)^2(s+2m_N^2)}{3 s^3}.
\end{eqnarray}
One can find that the production rate has little to do with the CM energy when
$m_N\ll\sqrt{s}$.
With the subsequent decay of $N\to \nu\gamma$, the signature to 
search is thus a single photon final state with missing energy, 
which  exhibits a typical monophoton signature at 
$e^+e^-$ colliders. 
The sterile neutrino $N$ decays into a photon and a SM active neutrino 
through the dipole operator, with the decay rate given by
\begin{equation}
\Gamma_{N\to\nu\gamma}=\frac{|d|^2m_N^3}{4\pi}.
\end{equation}

To make sure that there exists visible photon in the final state, the subsequent decay of $N$ must occur inside the fiducial volume of the detector. 
The probability of the heavy neutrino to
decay radiatively in the fiducial volume after traveling a distance $l$ from
the primary vertex is given by
\begin{equation}
P_{dec}(l)=(1-e^{-l/l_{dec}}){\rm Br}(N\to\nu\gamma).
\end{equation}
$l_{dec}$ is  the decay length of $N$,
which scales as 
\begin{equation}
l_{dec}=c\tau\beta\gamma=\frac{4\pi}{|d|^2m_N^4}\sqrt{E_N^2-m_N^2},
\end{equation}
where $E_N$ is the energy of $N$, with $E_N=\frac{s+m_N^2}{2\sqrt{s}}$ in the
process $e^+e^-\to  N\bar\nu$.
It is assumed that the $N$ decay is dominated by $N\to\nu\gamma$, hence the branching fraction  ${\rm Br}(N\to\nu\gamma)\simeq 1 $ is taken in this work.

Then, the number of events in signal can be given as 
\begin{equation}
N_s=L\sigma(e^+e^-\to N\nu){\rm Br}(N\to\nu\gamma)\epsilon_{cuts}\epsilon_{det}P_{dec}(l_D),
\end{equation}
where $l_D$ is the detector length, $L$ is the luminosity of the monophoton data collected at electron colliders, $\epsilon_{cuts}$
and $\epsilon_{det}$ are the efficiencies of the kinematic cuts and detection for the final photon, respectively.
Since $N$ is usually produced on-shell and travels some distance before decaying,
we employ the narrow width approximation to derive the kinematic imformation
of the final state photon.
The $1-\cos\theta$ distribution is used for the photon from 
$N$ decay, where $\theta$ is the photon angle in the rest frame of $N$ \cite{Li:1981um,Masip:2012ke}.

In the search for monophoton signature at electron colliders, the backgrounds can be classified into two categories:  the irreducible background and the reducible background.  The irreducible background arises from the neutrino production associated with one photon in SM $e^+e^-\to\nu\bar\nu\gamma$.  The reducible background comes from a photon in the final state together with several other visible particles which cannot be detected due to limitations of the detector acceptance. We will discuss the reducible background later in details for each experiment, since it strongly depends on the detector performance.

\subsection{Belle \uppercase\expandafter{\romannumeral2}}\label{sec:belle2}

At Belle II, photons and electrons are able to be detected in the 
Electromagnetic Calorimeter (ECL), which covers a polar angle region of $(12.4-155.1)^{\circ}$
and has inefficient gaps between the endcaps and the barrel for polar angles
between $(31.4-32.2)^{\circ}$ and $(128.7-130.7)^{\circ}$ in the lab frame  \cite{Kou:2018nap}.
The reducible background for monophoton searches at Belle II comes from 
two major parts:
one is mainly owing to the lack of polar angle  coverage 
of the ECL near the beam direction, which is referred to 
as the ``bBG''; 
the other one is mainly due to the gaps between 
the three segments in the ECL detector,  
which is referred to as the ``gBG''\cite{Kou:2018nap}. 
The bBG  arises from the electromagnetic processes $e^+e^-\to \gamma +\slashed{X}$
{with $\slashed{X}$ denoting the other particle (or particles) in the final state that are undetected due to the limitations of the detectors}
,
dominated by $e^+e^-\to\gamma\slashed{\gamma}(\slashed{\gamma})$ and $e^+e^-\to\gamma\slashed{e}^+\slashed{e}^-$,
where except the detected photon all the other final state particles 
are emitted along the beam directions with 
$\theta>155.1^{\circ}$ or $\theta<12.4^{\circ}$ in the lab frame. 
At Belle II, we apply the detector cuts for the final detected photon 
(hereafter the ``{\it basic cuts}"):
$12.4^{\circ}<\theta_\gamma<155.1^{\circ}$ in the lab frame.

For the asymmetric Belle II detector, 
in the bBG the maximum energy of the monophoton events 
in the CM frame,
$E_\gamma^m$,  is given by \cite{Liang:2019zkb, Zhang:2020fiu}
(if not exceeding $\sqrt{s}/2$) 
\begin{equation}
E_\gamma^m(\theta_\gamma) = 
\frac{ \sqrt{s}(A\cos\theta_1-\sin\theta_1)}
{A(\cos\theta_1-\cos\theta_\gamma)-(\sin\theta_\gamma+\sin\theta_1)},
\label{eq:bBG}
\end{equation}
where  
all angles are given in the CM frame, 
and $A=(\sin\theta_1-\sin\theta_2)/(\cos\theta_1-\cos\theta_2)$, 
with $\theta_1$ and $\theta_2$ being 
the polar angles corresponding to 
the edges of the ECL detector.
To remove the above bBG, we adopt the detector cut 
\begin{equation}
E_\gamma > E_\gamma^m  
\end{equation}
for the final monophoton (hereafter the {\it``bBG cut"}).

The monophoton energy in the gBG can be quite large in the central 
$\theta_{\gamma}$ region, since the gaps in the  ECL are significantly away from the beam direction.
The gBG have been simulated by 
Ref. \cite{Kou:2018nap} to search for an  
invisibly decaying dark photon.
In Ref. \cite{Kou:2018nap}, two different sets of detector cuts
are designed to optimize the detection efficiency for different masses of
the dark photon:
the {\it``low-mass cut"} and {\it``high-mass cut"}.
The {\it``low-mass cut"} can be described as $\theta_{\rm min}^{\rm low}
< \theta_{\gamma}^{\rm lab}<\theta_{\rm max}^{\rm low}$, where 
$\theta_{\rm min}^{\rm low}$ and $\theta_{\rm max}^{\rm low}$
are the minimum and maximum angles for the photon in the lab frame and can be respectively
fitted as functions of \cite{Duerr:2019dmv}
\begin{eqnarray}
	\theta_{\mathrm{min}}^{\mathrm{low}} &=& 5.399^{\circ} E_{\mathrm{CM}}(\gamma)^{2} / \mathrm{GeV}^{2}-58.82^{\circ} E_{\mathrm{CM}}(\gamma) / \mathrm{GeV}+195.71^{\circ}, \\
	\theta_{\mathrm{max}}^{\mathrm{low}}  &=& -7.982^{\circ} E_{\mathrm{CM}}(\gamma)^{2} / \mathrm{GeV}^{2}+87.77^{\circ} E_{\mathrm{CM}}(\gamma) / \mathrm{GeV}-120.6^{\circ},
\end{eqnarray}
with $E_{\mathrm{CM}}$ being the photon energy in the CM frame.
The {\it``high-mass cut"} can be described as $\theta_{\rm min}^{\rm high}
< \theta_{\gamma}^{\rm high}<\theta_{\rm max}^{\rm high}$, where 
$\theta_{\rm min}^{\rm high}$ and $\theta_{\rm max}^{\rm high}$
can be respectively
fitted as functions of \cite{Duerr:2019dmv}
\begin{eqnarray}
\theta_{\mathrm{min}}^{\mathrm{low}} &=& 3.3133^{\circ} E_{\mathrm{CM}}(\gamma)^{2} / \mathrm{GeV}^{2}-33.58^{\circ} E_{\mathrm{CM}}(\gamma) / \mathrm{GeV}+108.79^{\circ}, \\
\theta_{\mathrm{max}}^{\mathrm{low}}  &=& -5.9133^{\circ} E_{\mathrm{CM}}(\gamma)^{2} / \mathrm{GeV}^{2}+54.119^{\circ} E_{\mathrm{CM}}(\gamma) / \mathrm{GeV}-13.781^{\circ}.
\end{eqnarray}

\begin{figure}[htbp]
	\begin{centering}
		\includegraphics[width=0.8\columnwidth]{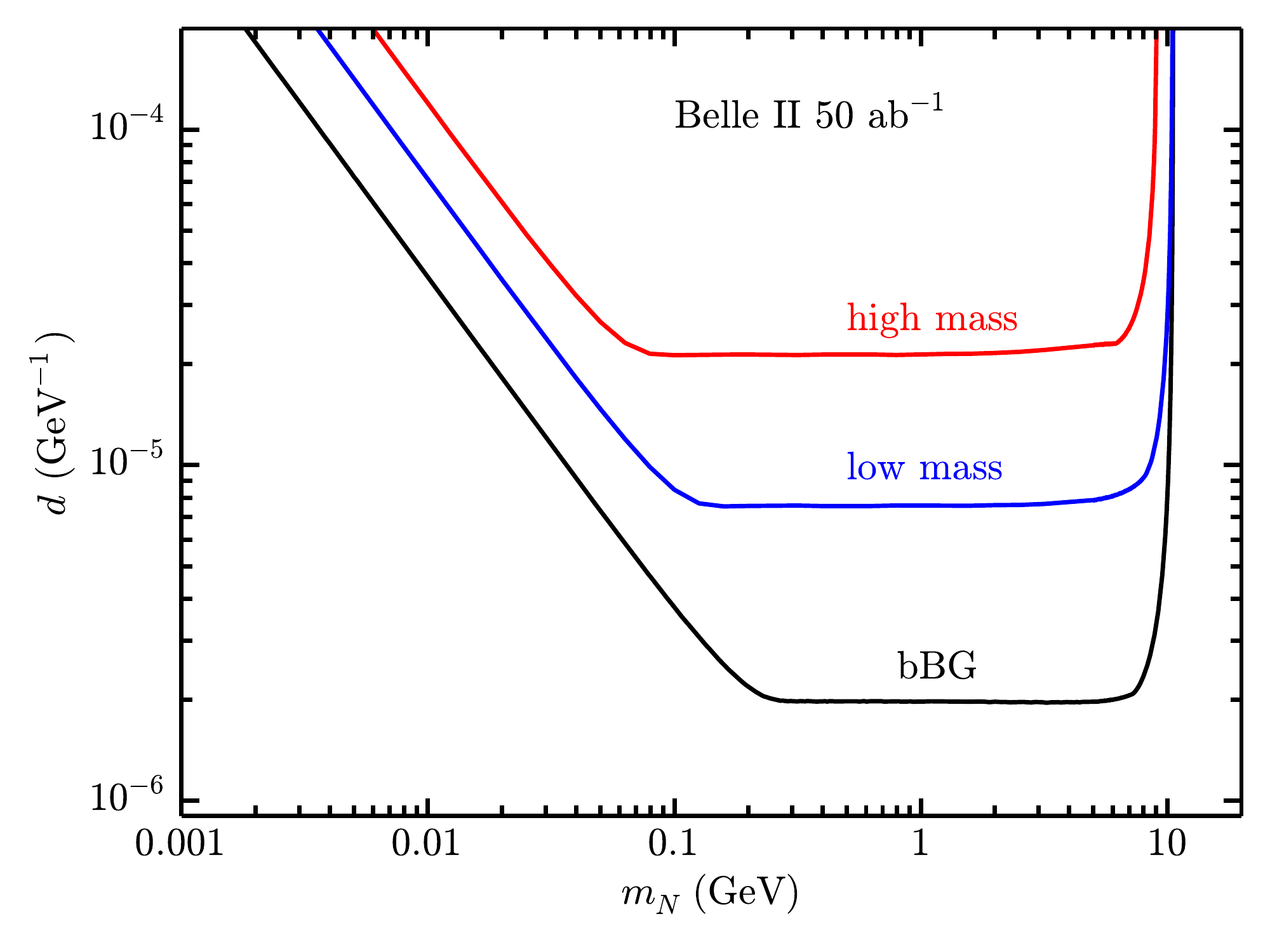}
		\caption{The expected 95\% C.L. upper limit on neutrino dipole coupling $d$ 
		at Belle II under {\it``low-mass cut"} (blue line), {\it``high-mass cut"} (red line ) and
	{\it``bBG cut"} (black line), with 50 ab$^{-1}$ integrated luminosity.}
		\label{fig:belle2}
	\end{centering}
\end{figure}

To probe the sensitivity for active-sterile neutrino transition magnetic moments $d$, 
we define $\chi^2(d)\equiv S^2/(S+B)$, where $S$ ($B$) is the number of events in the signal (background) processes. For background, $B=B_{\rm ir}+B_{\rm re}$ consists of the  number of events 
in irreducible background $B_{\rm ir}$ and reducible background $B_{\rm re}$.
By solving $\chi^2(d_{95})-\chi^2(0)=2.71$, one can obtain the 95\% confidence level (C.L.) 
upper limit for the neutrino dipole coupling $d_{95}$.
Fig. \ref{fig:belle2} shows the limits under  {\it``low-mass cut"} and {\it``high-mass cut"} with 50 ab$^{-1}$ integrated luminosity.
The number of events in irreducible background {is} calculated 
by integrating the differential cross sections under different detector cuts.
In addition, the detection efficiency of photon $\epsilon_{\rm dec}$
is assumed as 95\% \cite{Kou:2018nap}.
Following {Refs.} \cite{Belle:2000cnh, Dolan:2017osp}, we take the detector length $l_D=3$ m for Belle II.
For the reducible background, it is found that about 300 (25000) gBG events
survived the {\it``low-mass cut"} ({\it``high-mass cut"})  with 20 fb$^{-1}$ integrated luminosity \cite{Kou:2018nap}, which are rescaled according
to the considered luminosity. We can see that the constraint with the 
{\it``low-mass cut"} is always better than the {\it``high-mass cut"} in the whole
plotted mass region of $N$ at Belle II.

In order to compare with other experiments where  detailed simulations with gBG are not available, we also investigate the limits without taking 
gBG into account. In this scenario, the reducible background
can be removed with the {\it``bBG cut"}, and now the background events all
come from the irreducible backgrounds that survived the {\it``bBG cut"}.
The 95\% C.L. upper bound on $d$ under the {\it``bBG cut"} with gBG omitted is shown in Fig. \ref{fig:belle2}. It can be found that the upper bound under the {\it``bBG cut"} is about 4 times stronger {than} the one when gBG is considered 
under the {\it``low-mass cut"} with $0.3\ {\rm GeV} \lesssim m_N \lesssim 9$ GeV.

\subsection{BESIII and STCF}

At BESIII and STCF, we follow the cuts for photons applied by BESIII Collaboration  (hereafter the ``basic cuts"):
$E_\gamma > 25\ \MeV$ in the barrel ($|z_\gamma|<0.8$)  or 
$E_\gamma > 50\ \MeV$ in the end-caps ($0.86<|z_\gamma|<0.92$) \cite{Ablikim:2017ixv}.
In the following, we use the BESIII detector parameters to analyze the
constraints from STCF since {these} two experiments are similar. 
BESIII has not released any analysis
on gBG to our knowledge, thus we neglect gBG in the monophoton reducible background at BESIII and STCF. In the BESIII and STCF analyses, 
the monophoton reducible background at the electron-positron
colliders can be removed by applying the detector cut \cite{Liu:2019ogn}:
\begin{equation}
E_\gamma >E_b(\theta_\gamma)= \frac{\sqrt{s}}{(1+{\sin\theta_\gamma}/{\sin\theta_b})},
\label{eq:adv-cuts}
\end{equation}
on the final state photon,
where $E_b$ is the maximum energy of the photon with the polar angle $\theta_b$, and   $\theta_b$ denotes
the angle at the boundary of the sub-detectors.
Taking into account the coverage of main drift
chamber (MDC), electromagnetic calorimeter (EMC), and
time-of-flight (TOF), we have the polar angel $\cos{\theta_b}=0.95$ at BESIII \cite{Liu:2018jdi}.
In the following, we collectively refer to the ``basic cuts" and cut (\ref{eq:adv-cuts}) as the ``advanced cuts".

Since photon reconstruction
efficiencies are all more than 99\% \cite{BESIII:2011ysp} at BESIII, we
take $\epsilon_{det}\simeq 100\%$ for the photon at BESIII and STCF in this work.
The characteristic length scale of the detector is taken to be 80 cm,
which is the outer radius of MDC at BESIII . Thus, we set the detector length $l_D=0.8$ m for BESIII and STCF.

\begin{figure}[htbp]
	\begin{centering}
		\includegraphics[width=0.8\columnwidth]{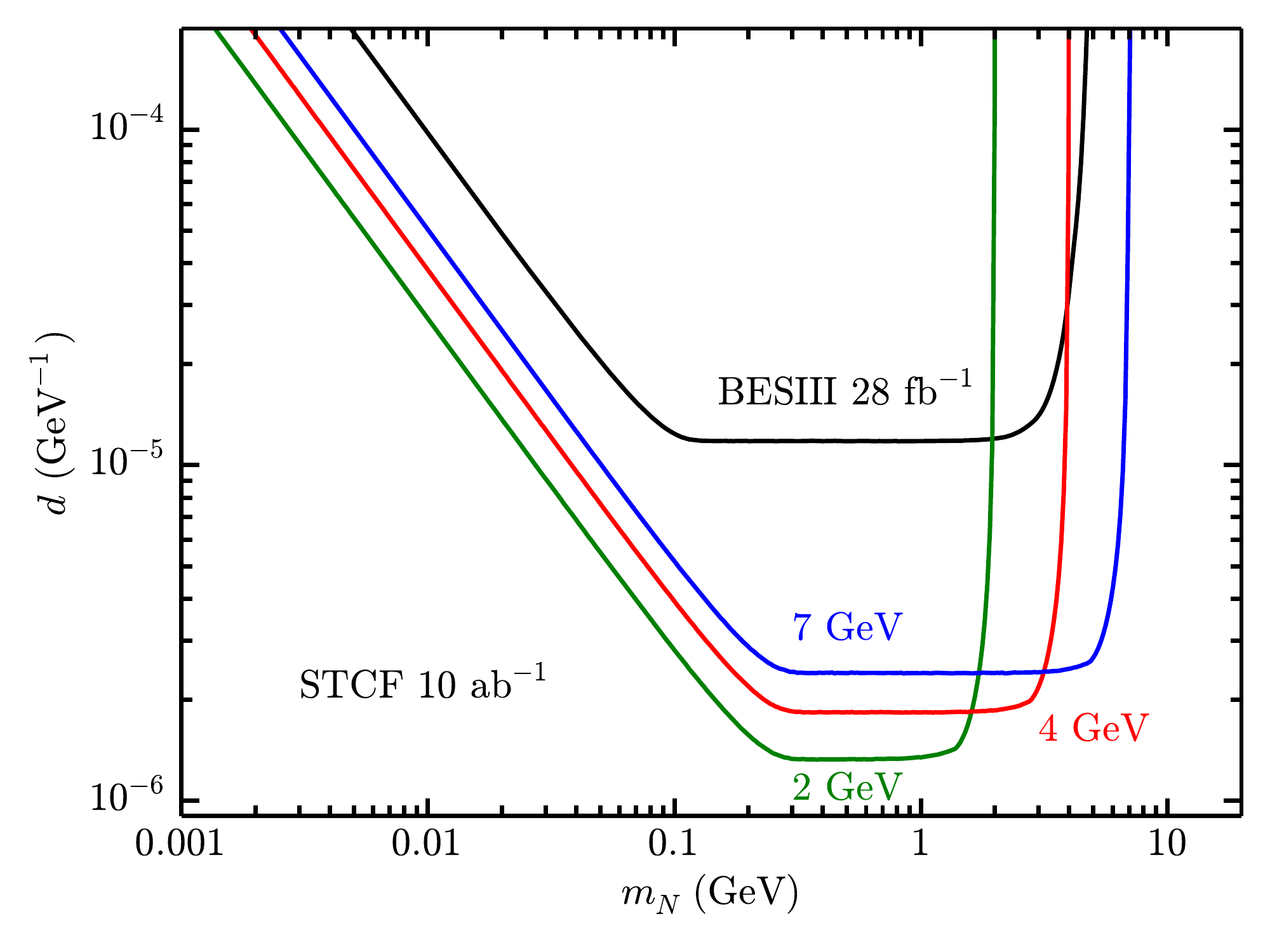}
		\caption{The expected 95\% C.L. upper bound on neutrino dipole coupling $d$ at BESIII and STCF under "advanced cut".
		The limits from BESIII (black line) is obtained with about 28 fb$^{-1}$ integrated luminosity collected at the various CM eneryies
	from 2.125 GeV to 4.95 GeV since 2012. The expected limits from STCF are shown at three typical energy points with 10 ab$^{-1}$ integrated luminosity for
$\sqrt{s}$= 2 (green line), 4 (red line), and 7 GeV (blue line), respectively. }
		\label{fig:bes3}
	\end{centering}
\end{figure}

At BESIII the monophoton trigger has been
implemented since 2012, and until now the corresponding events have been collected with the luminosity of about  28 fb$^{-1}$ at the CM energy from 2.125 GeV to 4.95 GeV \cite{bes3-data}.
We compute the number of events due to signal ($S$) and backgrounds ($B$) under the 
``advanced cuts", and define $\chi_{\rm tot}^2(d)=\sum_i\chi_{i}^2(d)$,
where $\chi_{i}^2(d)\equiv S_i^2/(S_i+B_i)$ for each BESIII colliding energy.
We show the expected  95\% C.L. upper limit on $d$ with 28 fb $^{-1}$ luminosity
collected at BESIII in Figs. \ref{fig:bes3} and \ref{fig:total}, which is obtained by demanding $\chi_{\rm tot}^2(d_{95})= \chi^2(0)+2.71$.
Fig. \ref{fig:bes3} also shows the expected limits on $d$ with assumed 10 ab$^{-1}$ luminosity 
at three different colliding energies in future STCF, respectively.
One can find that, with same luminosity, operated at lower
energy, STCF has better sensitivity  in probing the low-mass region.
This is because the monophoton cross section in small mass $N$ production is not too dependent on the CM energy, while decreases in the background with the increment of the CM energy.

\section{Results}
\label{sec:res}

\begin{figure}[htbp]
	\begin{centering}
		\includegraphics[width=0.8\columnwidth]{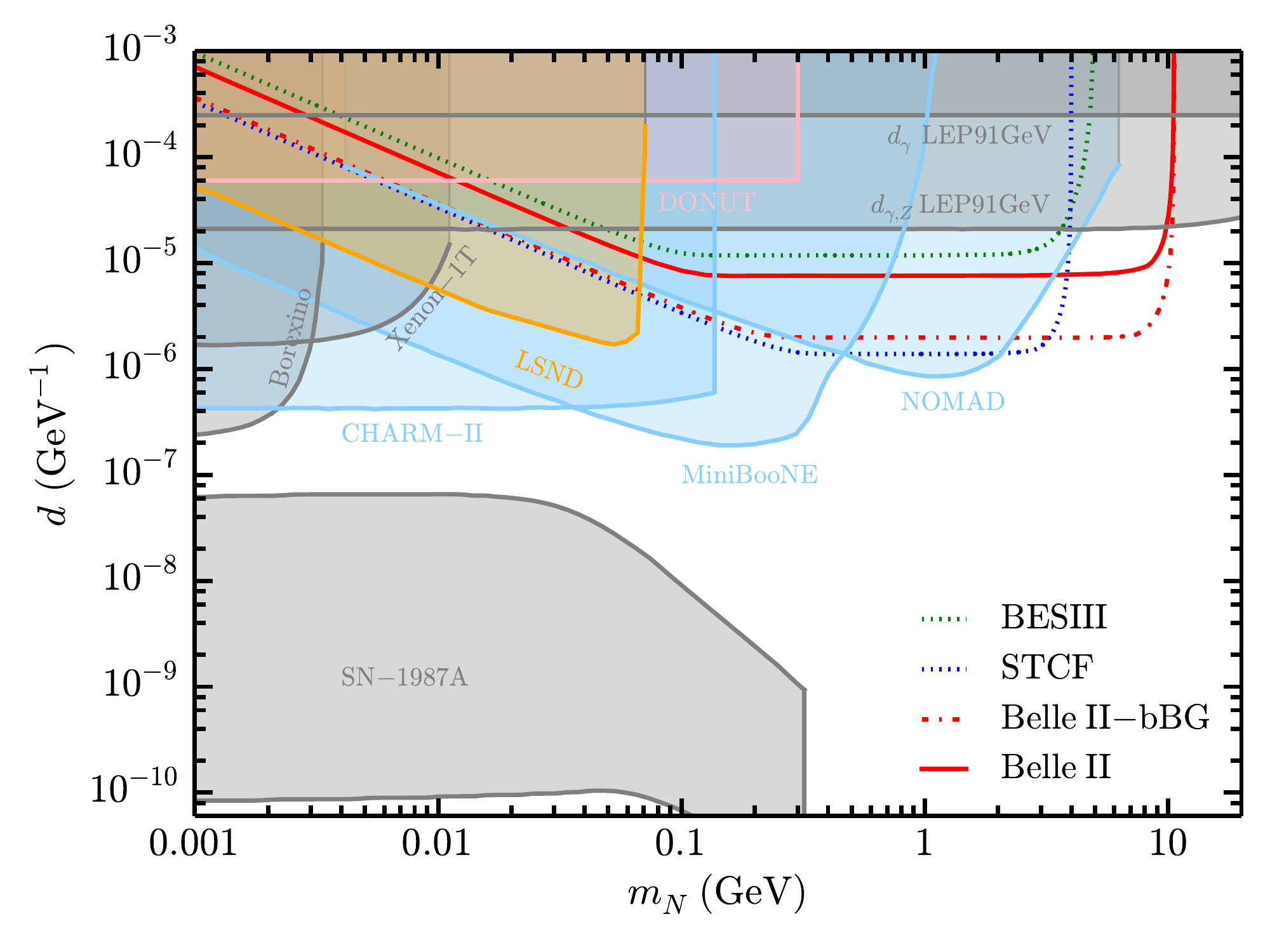}
		\caption{
			The expected 95\% C.L. exclusion limits on active-sterile neutrino transition magnetic moment $d$  at
			Belle II, BESIII, and STCF, which work for all the three SM neutrinos.
			The Belle II limit (red solid line) is obtained under the
			``{\it low-mass cut}", where both the bBG and
			the gBG are considered. The other Belle II limit (red dot-dashed line) is obtained with the ``{\it bBG cut}" where the gBG is omitted.
			The BESIII limit
			(green dotted line) is obtained with the integrated luminosity of 28 fb$^{-1}$ during 2012-2020 where the gBG is omitted.
			The STCF limits
			(blue dotted line) is obtained for $\sqrt{s}=$ 4 GeV and 30 ab$^{-1}$ under the ``advanced cuts", where
			the gBG is omitted.
			The landscape of current constraints are also presented with shaded region.
			The gray shaded regions  expliting the constraints from Borexino \cite{Brdar:2020quo}, Xenon1T \cite{Brdar:2020quo}, LEP \cite{Magill:2018jla}, and SN-1987A \cite{Magill:2018jla} are relevant for all three SM neutrinos. 
			The orange regions show the constraint on electron-neutrino dipole coupling $d_e$ from LSND \cite{Magill:2018jla}.
			The skyblue regions show the constraints on muon-neutrino dipole coupling $d_\mu$ from CHARM-II \cite{Coloma:2017ppo}, MiniBooNE \cite{Magill:2018jla}, and NOMAD \cite{Magill:2018jla,Gninenko:1998nn}. 
			The pink region shows the constraints on tau-neutrino dipole coupling $d_\tau$ from DONUT \cite{DONUT:2001zvi}.
		}
		\label{fig:total}
	\end{centering}
\end{figure}

The landscape of current constraints on transition magnetic moments are summarized in Fig. \ref{fig:total} with shaded region,
coming from terrestrial experiments such as  Borexino \cite{Brdar:2020quo}, Xenon-1T \cite{Brdar:2020quo}, CHARM-II \cite{Coloma:2017ppo}, MiniBooNE \cite{Magill:2018jla}, LSND \cite{Magill:2018jla}, NOMAD \cite{Gninenko:1998nn,Magill:2018jla},  DONUT \cite{DONUT:2001zvi}, and LEP \cite{Magill:2018jla}, and  astrophysics supernovae SN 1987A \cite{Magill:2018jla}. 
It is noted that constraints on transition magnetic moments involving
three SM active neutrinos ($\nu_{e,\mu,\tau}$)  are similar in XENON-1T,
Borexino, SN 1987A, and  LEP \cite{Brdar:2020quo, Magill:2018jla}, which are shaded in gray.

For $d_e$, the constraint can also be from LSND \cite{Magill:2018jla}, which  is shown in orange region.
The limits from CHARM-II \cite{Coloma:2017ppo}, MiniBooNE \cite{Magill:2018jla}, and NOMAD \cite{Magill:2018jla} are relevant only for $d_\mu$, which {are} shown in skyblue {regions}. The constraint
for $d_\mu$  also can be provided by LSND  \cite{Magill:2018jla}. Since it is not competitive with CHARM-II, it is not shown here. 
DONUT \cite{DONUT:2001zvi} gave an upper 90\% C.L. limit
on $d_\tau$ of 5.8 $\times 10^{-5}\ {\rm GeV}^{-1}$ for $m_N<0.3$ GeV , which is shown with pink curve.

In order to describe more intuitively, we summarize the sensitivity
on $d$ at 95\% C.L. from the low-energy electron colliders, including Belle II,
BESIII and STCF, which works for all the three neutrinos.
The Belle II limit shown as solid curve in Fig. \ref{fig:total},
has been analyzed taking into account the various SM backgrounds.
With 50 ab$^{-1}$ data, transition magnetic moment down to about 8 $\times 10^{-6}\ {\rm GeV}^{-1}$ for mass about (0.1-7) GeV is expected to be probed by Belle II,
which has better sensitivity than LEP \cite{Magill:2018jla}.
A larger parameter space for $d_e$ and $d_\tau$ in the mass region of 0.07-10 GeV 
and for $d_\mu$ in the mass region of 2-10 GeV is
previously unconstrained by other experiments will be explored
by Belle II.
And for mass from about 3 GeV to 10 GeV, Belle II can still provide the
most stringent restrictions on $d_\mu$.

The STCF and BESIII limits,  shown as  dotted curves in
Fig. \ref{fig:total}, are obtained when the background due to the gaps in 
detectors is neglected.
Using the monophoton signature  with about 28 fb$^{-1}$ luminosity collected at the CM energy from 2.125 GeV to 4.95 GeV during 2012-2020, the BESIII can probe the upper limit of $d$ down to about 1.3 $\times 10^{-5}\ {\rm GeV}^{-1}$ for $0.1\ {\rm GeV}<m_N<3\ {\rm GeV}$, which is still competitive with LEP \cite{Magill:2018jla}, and can provide better sensitivity for $d_e$ and $d_\tau$
than exsiting experiments.
With 30 ab$^{-1}$ data at $\sqrt{s}=4$ GeV, STCF can provide constraints on  
transition magnetic moments $d\lesssim 1.3 \times 10^{-6}\ {\rm GeV}^{-1}$ for
mass from 0.2 GeV to about 3 GeV. The expected limit from STCF running at 4 GeV
with  30 ab$^{-1}$ data  can also given leading sensitivity on  $d_e$ and $d_\tau$ 
for mass $m_N$ $\sim$ (0.07-3.8) GeV and on $d_\mu$ for $m_N$ $\sim$ (2-3.8) GeV.

The omission of the gBG in BESIII (28 fb$^{-1}$) leads to an almost comparable limit
with Belle II (50 ab$^{-1}$) with gBG included for $m_N\lesssim 3$ GeV.
In order to compare the capability of probing
the parameter space from different experiments, we also
present a Belle II limit (dot-dashed curve) without gBG
considered. It can be found that the luminosity of STCF is lower than
Belle II, but STCF has better sensitivity in probing the low-mass region
($m_N\lesssim$ 3.5 GeV) than Belle II. 
This is because
STCF is operated at a lower colliding energy where the
monophoton cross section in SM is smaller than Belle II. 
About four times of magnitude difference 
in sensitivity between the two Belle II limits, the solid curve
and the dot-dashed curve in Fig. \ref{fig:total} show that the control
on gBG is important in probing the neutrino dipole portal.

\section{Summary}
\label{sec:sum}
In this paper, we analyzed the sensitivity on active-sterile neutrino transition magnetic moment $d$ from several electron colliders
operated at few GeV: Belle II, BESIII and STCF. With about 28 fb$^{-1}$ 
luminosity in monophoton data collected during 2012-2020, the BESIII 
can provide the expected upper limit of $d$ down to 1.3 $\times 10^{-5}\ {\rm GeV}^{-1}$ for $0.1\ {\rm GeV}<m_N<3\ {\rm GeV}$. Projected limits with Belle II
and STCF experiments are also analyzed. Belle II with 50 ab$^{-1}$ monophoton data
can probe transition magnetic moment down to 8 $\times 10^{-6}\ {\rm GeV}^{-1}$ for mass about (0.1-7) GeV, and will improve the sensitivity about four times with
gBG omitted. The future 4 GeV STCF with 30 ab$^{-1}$ monophoton data can further improve the sensitivity to low-mass sterile neutrino ($m_N\lesssim$ 3.5 GeV) than Belle II.
In general, BESIII, STCF and Belle II can explore the parameter space
for $d_e$ and $d_\tau$, which is previously unconstrained by other exsiting
experiments. And Belle II and STCF can also provide leading constraints on 
$d_\mu$ for high-mass sterile neutrino.

\acknowledgments
This work was supported in part by the National Natural Science Foundation of China (Grants No.12105327, No. 11805001) and 
the Key Research Foundation of Education Ministry of Anhui Province of China (No.KJ2021A0061).


\begin{thebibliography}{999}
\bibitem{Magill:2018jla}
G.~Magill, R.~Plestid, M.~Pospelov and Y.~D.~Tsai,
Phys. Rev. D \textbf{98} (2018) no.11, 115015
doi:10.1103/PhysRevD.98.115015
[arXiv:1803.03262 [hep-ph]].

\bibitem{Schwetz:2020xra}
T.~Schwetz, A.~Zhou and J.~Y.~Zhu,
JHEP \textbf{21} (2020), 200
doi:10.1007/JHEP07(2021)200
[arXiv:2105.09699 [hep-ph]].

\bibitem{Borexino:2017fbd}
M.~Agostini \textit{et al.} [Borexino],
Phys. Rev. D \textbf{96} (2017) no.9, 091103
doi:10.1103/PhysRevD.96.091103
[arXiv:1707.09355 [hep-ex]].

\bibitem{Brdar:2020quo}
V.~Brdar, A.~Greljo, J.~Kopp and T.~Opferkuch,
JCAP \textbf{01} (2021), 039
doi:10.1088/1475-7516/2021/01/039
[arXiv:2007.15563 [hep-ph]].

\bibitem{XENON:2020rca}
E.~Aprile \textit{et al.} [XENON],
Phys. Rev. D \textbf{102} (2020) no.7, 072004
doi:10.1103/PhysRevD.102.072004
[arXiv:2006.09721 [hep-ex]].

\bibitem{CHARM-II:1989srx}
D.~Geiregat \textit{et al.} [CHARM-II],
Phys. Lett. B \textbf{232} (1989), 539
doi:10.1016/0370-2693(89)90457-7

\bibitem{Coloma:2017ppo}
P.~Coloma, P.~A.~N.~Machado, I.~Martinez-Soler and I.~M.~Shoemaker,
Phys. Rev. Lett. \textbf{119} (2017) no.20, 201804
doi:10.1103/PhysRevLett.119.201804
[arXiv:1707.08573 [hep-ph]].

\bibitem{MiniBooNE:2007uho}
A.~A.~Aguilar-Arevalo \textit{et al.} [MiniBooNE],
Phys. Rev. Lett. \textbf{98} (2007), 231801
doi:10.1103/PhysRevLett.98.231801
[arXiv:0704.1500 [hep-ex]].

\bibitem{LSND:1996ubh}
C.~Athanassopoulos \textit{et al.} [LSND],
Phys. Rev. Lett. \textbf{77} (1996), 3082-3085
doi:10.1103/PhysRevLett.77.3082
[arXiv:nucl-ex/9605003 [nucl-ex]].

\bibitem{Gninenko:1998nn}
S.~N.~Gninenko and N.~V.~Krasnikov,
Phys. Lett. B \textbf{450} (1999), 165-172
doi:10.1016/S0370-2693(99)00130-6
[arXiv:hep-ph/9808370 [hep-ph]].

\bibitem{NOMAD:1998pxi}
J.~Altegoer \textit{et al.} [NOMAD],
Phys. Lett. B \textbf{428} (1998), 197-205
doi:10.1016/S0370-2693(98)00402-X
[arXiv:hep-ex/9804003 [hep-ex]].

\bibitem{DONUT:2001zvi}
R.~Schwienhorst \textit{et al.} [DONUT],
Phys. Lett. B \textbf{513} (2001), 23-29
doi:10.1016/S0370-2693(01)00746-8
[arXiv:hep-ex/0102026 [hep-ex]].

\bibitem{Plestid:2020vqf}
R.~Plestid,
doi:10.1103/PhysRevD.104.075027
[arXiv:2010.04193 [hep-ph]].

\bibitem{Jodlowski:2020vhr}
K.~Jod\l{}owski and S.~Trojanowski,
JHEP \textbf{05} (2021), 191
doi:10.1007/JHEP05(2021)191
[arXiv:2011.04751 [hep-ph]].

\bibitem{Ismail:2021dyp}
A.~Ismail, S.~Jana and R.~M.~Abraham,
[arXiv:2109.05032 [hep-ph]].

\bibitem{Shoemaker:2018vii}
I.~M.~Shoemaker and J.~Wyenberg,
Phys. Rev. D \textbf{99} (2019) no.7, 075010
doi:10.1103/PhysRevD.99.075010
[arXiv:1811.12435 [hep-ph]].

\bibitem{Miranda:2021kre}
O.~G.~Miranda, D.~K.~Papoulias, O.~Sanders, M.~T\'ortola and J.~W.~F.~Valle,
[arXiv:2109.09545 [hep-ph]].

\bibitem{Asner:2008nq}
D.~M.~Asner, T.~Barnes, J.~M.~Bian, I.~I.~Bigi, N.~Brambilla, I.~R.~Boyko, V.~Bytev, K.~T.~Chao, J.~Charles and H.~X.~Chen, \textit{et al.}
Int. J. Mod. Phys. A \textbf{24} (2009), S1-794
[arXiv:0809.1869 [hep-ex]].

\bibitem{Belle-II:2010dht}
T.~Abe \textit{et al.} [Belle-II],
[arXiv:1011.0352 [physics.ins-det]].

\bibitem{Luo:2019xqt}
Q.~Luo, W.~Gao, J.~Lan, W.~Li and D.~Xu,
doi:10.18429/JACoW-IPAC2019-MOPRB031

\bibitem{Charm-TauFactory:2013cnj}
A.~E.~Bondar \textit{et al.} [Charm-Tau Factory],
Phys. Atom. Nucl. \textbf{76} (2013), 1072-1085
doi:10.1134/S1063778813090032

\bibitem{Shi:2020nrf}
X.~D.~Shi, X.~R.~Zhou, X.~S.~Qin and H.~P.~Peng,
JINST \textbf{16} (2021) no.03, P03029
doi:10.1088/1748-0221/16/03/P03029
[arXiv:2011.01654 [physics.ins-det]].

\bibitem{Li:1981um}
L.~F.~Li and F.~Wilczek,
Phys. Rev. D \textbf{25} (1982), 143
doi:10.1103/PhysRevD.25.143

\bibitem{Masip:2012ke}
M.~Masip, P.~Masjuan and D.~Meloni,
JHEP \textbf{01} (2013), 106
doi:10.1007/JHEP01(2013)106
[arXiv:1210.1519 [hep-ph]].

\bibitem{Kou:2018nap}
E.~Kou \textit{et al.} [Belle-II],
PTEP \textbf{2019} (2019) no.12, 123C01
[erratum: PTEP \textbf{2020} (2020) no.2, 029201]
doi:10.1093/ptep/ptz106
[arXiv:1808.10567 [hep-ex]].

\bibitem{Duerr:2019dmv}
M.~Duerr, T.~Ferber, C.~Hearty, F.~Kahlhoefer, K.~Schmidt-Hoberg and P.~Tunney,
JHEP \textbf{02} (2020), 039
doi:10.1007/JHEP02(2020)039
[arXiv:1911.03176 [hep-ph]].

\bibitem{Liang:2019zkb}
J.~Liang, Z.~Liu, Y.~Ma and Y.~Zhang,
Phys. Rev. D \textbf{102} (2020) no.1, 015002
doi:10.1103/PhysRevD.102.015002
[arXiv:1909.06847 [hep-ph]].

\bibitem{Zhang:2020fiu}
Y.~Zhang, Z.~Yu, Q.~Yang, M.~Song, G.~Li and R.~Ding,
Phys. Rev. D \textbf{103} (2021) no.1, 015008
doi:10.1103/PhysRevD.103.015008
[arXiv:2012.10893 [hep-ph]].

\bibitem{Belle:2000cnh}
A.~Abashian \textit{et al.} [Belle],
Nucl. Instrum. Meth. A \textbf{479} (2002), 117-232
doi:10.1016/S0168-9002(01)02013-7

\bibitem{Dolan:2017osp}
M.~J.~Dolan, T.~Ferber, C.~Hearty, F.~Kahlhoefer and K.~Schmidt-Hoberg,
JHEP \textbf{12} (2017), 094
[erratum: JHEP \textbf{03} (2021), 190]
doi:10.1007/JHEP12(2017)094
[arXiv:1709.00009 [hep-ph]].

\bibitem{Ablikim:2017ixv}
M.~Ablikim \textit{et al.} [BESIII],
Phys. Rev. D \textbf{96} (2017) no.11, 112008
doi:10.1103/PhysRevD.96.112008
[arXiv:1707.05178 [hep-ex]].

\bibitem{Liu:2019ogn}
Z.~Liu, Y.~H.~Xu and Y.~Zhang,
JHEP \textbf{06} (2019), 009
doi:10.1007/JHEP06(2019)009
[arXiv:1903.12114 [hep-ph]].

\bibitem{Liu:2018jdi}
Z.~Liu and Y.~Zhang,
Phys. Rev. D \textbf{99} (2019) no.1, 015004
doi:10.1103/PhysRevD.99.015004
[arXiv:1808.00983 [hep-ph]].

\bibitem{BESIII:2011ysp}
M.~Ablikim \textit{et al.} [BESIII],
Phys. Rev. D \textbf{83} (2011), 112005
doi:10.1103/PhysRevD.83.112005
[arXiv:1103.5564 [hep-ex]].

\bibitem{bes3-data}
http://english.ihep.cas.cn/bes/doc/2250.html

\end{thebibliography}
\end{document}